# Direction-Controlled Chemical Doping for Reversible G-Phonon Mixing in ABC Trilayer Graphene

Kwanghee Park[1] and Sunmin Ryu[2]*

[1]Department of Applied Chemistry, Kyung Hee University, Yongin, Gyeonggi 446-701, Korea
[2]Department of Chemistry, Pohang University of Science and Technology (POSTECH), Pohang, Gyeongbuk 790-784, Korea

*E-mail: sunryu@postech.ac.kr

**Abstract**

Not only the apparent atomic arrangement but the charge distribution also defines the crystalline symmetry that dictates the electronic and vibrational structures. In this work, we report reversible and direction-controlled chemical doping that modifies the inversion symmetry of AB-bilayer and ABC-trilayer graphene. For the "top-down" and "bottom-up" hole injection into graphene sheets, we employed molecular adsorption of electronegative $I_2$ and annealing-induced interfacial hole doping, respectively. The chemical breakdown of the inversion symmetry led to the mixing of the G phonons, Raman active $E_g$ and Raman-inactive $E_u$ modes, which was manifested as the two split G peaks, $G^-$ and $G^+$. The broken inversion symmetry could be recovered by removing the hole dopants by simple rinsing or interfacial molecular replacement. Alternatively, the symmetry could be regained by double-side charge injection, which eliminated $G^-$ and formed an additional peak, $G^o$, originating from the barely doped interior layer. Chemical modification of crystalline symmetry as demonstrated in the current study can be applied to other low dimensional crystals in tuning their various material properties.

**Introduction**

Many layered crystals such as graphite, hexagonal BN and 2H-$MoS_2$ consist of atom-thick two-dimensional building blocks, which have been studied by numerous research groups in the past decade.[1-13] The interlayer bonding of the materials are mainly of van der Waals (vdW) type and typically two-orders of magnitude weaker than their intralayer counterparts.[14,15] However, it is the lack of the weak coupling that makes graphene distinct from graphite,[16] and thus few-layer graphene (FLG) with different number of layers exhibit significant difference in their physical and chemical properties.[16-19]

In addition, the crystallographic stacking order in FLG sheets serves as another degree of freedom that dictates the lattice symmetry affecting their thermodynamic stability, electronic, optical and magnetic properties. [20-24] In this regard, trilayer (3L) graphene, the thinnest crystal with multiple stable polytypes, has recently been drawing a great research interest.[22,23] Trilayer graphene consists of ~85% semimetallic Bernal-stacked 3L (ABA) and ~15% semiconducting rhombohedral 3L (ABC) domains, when mechanically exfoliated from kish graphite.[23] The dominance of the Bernal stacking agrees with its slightly larger cohesion energy[25] and is also found in natural bulk graphite.[26]

With the slight difference in their stacking order, the two polytypes belong to different space groups exhibiting distinctive lattice symmetries and phonon structures.[23,27,28] Due to the symmetry-imposed selections rules,[29] optical transitions can serve as powerful probes for the diverse structural landscape. Stacking graphene layers, in particular, may or may not conserve the inversion symmetry that dictates the Raman and IR activities of phonon modes. When stacked into AB bilayer, for example, the inversion symmetry in graphene is maintained and the even-parity $E_{2g}$ (G) mode of graphene evolves into even-parity $E_g$ and odd-parity $E_u$ modes, respectively corresponding to the in-phase and out-of-phase movements of the two layers.[27] An additional graphene layer induces further branching into $2E'$ + $E''$ for non-centrosymmetric ABA and $2E_g + E_u$ for centrosymmetric ABC.[27] Since parity is not conserved for ABA, all the three modes of ABA are of indefinite parity and found to be active for both Raman and IR transitions. In contrast, parity is still conserved quantity for ABC, and thus Raman and IR transitions are exclusively allowed for the even and odd-parity modes, respectively. [23, 27]

Due to the finite number of layers of FLG, its inversion symmetry and optical transitions can be readily modified by various external perturbation unlike bulk graphite, which should provide a new spectroscopic degree of freedom in understanding and controlling underlying lattice dynamics. Whereas the $E_u$ mode of AB is silent in the Raman scattering, it can be activated when the inversion symmetry of AB is broken by electrical gating, for example. Due to the selective coupling with the interband electronic transitions predicted by Ando et al.,[30] its frequency undergoes a downshift in contrast to the upshifting $E_g$ mode exhibiting the apparent G-peak splitting as decreasing the Fermi level ($E_F$).[31] It was further proposed that the broken symmetry mixes the two modes generating the two Raman-active superposed modes,[32,33] which was soon confirmed by electrical gating using polymer electrolytes.[34,35] Considering its intimate contact with underlying substrates[36] and facile molecular adsorption,[37] the lattice symmetry of FLG may also be controlled using various chemical perturbation. Although there have been a few reports on chemical modulation of lattice symmetry of 2L[38,39] and 4L,[38] the molecular nature of the dopants and thus doping mechanisms were not clearly identified. Moreover, it has not been explored experimentally how phonon excitation of different polytypes, ABA and ABC for example, would respond to such symmetry change. In this work, we demonstrate reversible "top-down" and "bottom-up" chemical hole doping to modify the symmetries of AB, ABA and ABC FLG supported on SiO$_2$ substrates, by exploiting adsorption of electronegative I$_2$ molecules[37] and intercalation of O$_2$/H$_2$O redox couples,[40] respectively. The degree of the perturbation could be monitored by the frequency difference between the superposed G modes. We further show that the non-centrosymmetric ABA does not exhibit the G-peak splitting unlike ABC or AB, and that the chemical modification can be undone by simple rinsing or interfacial molecular replacement. Alternatively, the broken inversion symmetry of AB or ABC could be recovered by double-sided charge injection, which lead to a new pair of G peaks. The reversible and direction-controlled chemical charge doping as demonstrated in this study can also be useful in modifying the electronic and vibrational transitions of other low dimensional crystals.

As shown in Fig. 1a, the overall Raman spectra of pristine ABA and ABC trilayers supported on SiO$_2$/Si substrates (see Methods for preparation of samples and measurements) are very similar to each other. Nonetheless, there are a few distinctive spectral features that can be used in characterizing their stacking order. First, the G peak for the C-C stretching mode exhibits ~1 cm$^{-1}$ downshift for ABC with respect to ABA, presumably due to the different phonon band structures.[23] The G-peak lineshape of ABC is slightly narrower than that of ABA, since the former has weaker electron-phonon coupling than the latter consequently having longer G-phonon lifetime.[41] In addition, the 2D peak shows a more significant difference in its lineshape for both polytypes as shown in Fig. 1a. According to the double-resonance (DR) scattering model,[13] an electron and a hole excited by a Raman excitation photon scatter with two D phonons with wave vectors that match the intervalley resonant electronic transitions. Consequently polytypes of different electronic band structures would exhibit different 2D spectra even if their phonon band structures are identical. Since trilayer graphene has three sets of valence and conduction bands, there can be as many as 15 scattering processes that satisfy the requirement for DR.[42] As shown in Supplementary Fig. S1, however, 2D peaks of both polytypes can be satisfactorily fitted with 6 Lorentzian functions due to non-distinguishability and insufficient scattering probability of some among the 15 processes.[23,42] Whereas these and other spectral features can be used to characterize stacking domains by Raman mapping, the frequency of G peak ($\omega_G$) and the effective linewidth of 2D peak ($\Gamma_{2D}^{eff}$) were chosen for the simplicity and shorter analysis time. As shown in the Raman maps (Figs. 1d ~ 1f), the trilayer sheet consists of the two domains that are indistinguishable in the optical micrograph (Fig. 1c) or the intensity map (Fig. 1d). The domain denoted ABC indeed shows ~1 cm$^{-1}$ smaller $\omega_G$ and ~10 cm$^{-1}$ larger $\Gamma_{2D}^{eff}$ than the ABA-domain in agreement with previous reports.[23,41] Although $\Gamma_{2D}^{eff}$ defined as a single Lorentzian linewidth does not represent the lineshape of 2D peak very well, it was proven highly efficient in differentiating the stacking domains of 3L.[23]

Figure 2 presents the Raman spectra of AB, ABA and ABC samples the symmetries of which were being progressively modified by the bottom-up hole injection using the annealing-induced chemical doping.[19,43,44] (see Methods for details.) When ambient O$_2$ intercalates through the annealed

graphene/SiO$_2$ interface,[40,44] it undergoes a redox reaction involving the O$_2$/H$_2$O couple with an electrochemical potential sufficiently lower than the Fermi level of graphene.[45,46] Thus, the interfacial reaction consumes electrons in graphene above and essentially leads to bottom-up hole doping with a maximum hole density of ~2x10$^{13}$ /cm$^2$ in 1L for the annealing temperature (T$_{anneal}$) of 600 °C.[19] As increasing T$_{anneal}$ or essentially the hole density, the G peak of AB upshifts from its intrinsic frequency of ~1581 cm$^{-1}$ to ~1598 cm$^{-1}$. At 400 °C, a shoulder at lower frequency appears and essentially develops into a separate peak at higher temperatures. Both peaks were fitted with a double Lorentzian function and the low and high frequency peaks were labelled respectively as G$^-$ and G$^+$ according to Yan et al.[34] Whereas a similar upshift and splitting can be seen for ABC, ABA shows only upshift without splitting. The difference between ABA and ABC in the response to the bottom-up hole doping can be seen more clearly in Fig. 3 which shows the variation of the peak frequencies as a function of T$_{anneal}$. As the degree of the doping increases, G (or G$^+$) increases in its peak frequency by 10 ~ 15 cm$^{-1}$ for the three FLG systems. For AB and ABC, the new G$^-$ peak gradually downshifts with increasing charge density. Interestingly, the splitting in $\omega_G$ ($\Delta\omega_G$) reaches ~15 cm$^{-1}$ at 700 °C for both of AB and ABC despite the slight difference in their frequencies. Whereas the phonon frequencies are also subject to the lattice deformation of native or thermally induced origin,[43] $\Delta\omega_G$ is not significantly influenced by such effects to a first order approximation.

The unique spectral difference found in the two 3L-polytypes can be explained by the phonon mixing induced by the symmetry breaking and their distinctive coupling between the nuclear and electronic degrees of freedom.[32,33] As explained above, the G modes of the centrosymmetric ABC lattice consist of one IR-active $E_u$ and two Raman-active $E_{g,a}$ and $E_{g,b}$ modes, the last two of which are predicted to be 8 cm$^{-1}$ separated apart.[27] In contrast, pristine ABC shows only one peak at ~1581 cm$^{-1}$ as shown in Fig. 2, which implies that the $E_{g,b}$ mode with a higher frequency has negligible contribution to the G peak. As the strong interlayer electric field induced by the bottom-up charge doping renders the top and the bottom graphene layers unequal, the inversion symmetry of ABC cannot be maintained. Consequently, the three phonon modes are no longer the normal modes of the non-centrosymmetric ABC. Instead, a new set of eigenmodes can be approximately formed by superposing the unperturbed normal modes, presumably becoming all Raman-active.[34] Since Fig. 2 reveals apparently two peaks for ABC, however, it is likely that one of the new eigenmodes also has a small scattering probability as seen in the pristine ABC. When the $E_{g,b}$ mode with negligible spectral contribution is excluded from the mixing, the system of ABC is essentially identical to that of AB. Thus, we assign the two split Raman peaks (denoted G$^-$ and G$^+$) of ABC in Fig. 2 to the superposition states of $E_{g,a}$ and $E_u$ modes as in AB.[34] The upshift (downshift) of G$^+$ (G$^-$) induced by the increased hole density or decreased $E_F$ is due to the diminishing (growing) interband transitions that renormalize the phonon's self-energy.[47,48]

In Fig. 4a, the mixed phonons were shown to be decoupled by reversing the bottom-up hole doping. To remove the interfacial hole dopants, we exploited the water intercalation recently demonstrated by Lee et al.[40] After obtaining the Raman spectra exhibiting the G peak splitting induced by annealing at 500 °C (Fig. 4a), the sample shown in Fig. 4c was submerged in distilled water for 7 days to induce complete intercalation of one bilayer of water[49,50] that replaces the hole dopants and undoes the annealing induced hole doping from the graphene edges to the center as illustrated in Fig. 4b.[40] The top Raman spectrum in Fig. 4a shows a single symmetric G peak at ~1581 cm$^{-1}$ and confirms that the superposed states have been decoupled due to the undoping that restored the inversion symmetry. It is also to be noted that the 2D peak has been almost completely recovered to its pristine state. Using Raman mapping, we were also able to show that the mixing and the decoupling occur reversibly throughout the sample (Fig. 4c and 4d) with clear ABA-ABC domain boundaries as revealed in the $\Gamma_{2D}^{eff}$-map of Fig. 4f~4h. Although the annealing changed the intensity and lineshape of the 2D peak (Fig. 4a), the $\Gamma_{2D}^{eff}$-maps in Fig. 4f and 4g indicates that the stacking domains remained almost intact throughout the thermal activation, which is consistent with the previous observation.[23] Furthermore, Fig. 4h clearly shows that the water-intercalation recovered the numerical values of $\Gamma_{2D}^{eff}$ across the sample. The $\Delta\omega_G$-map in Fig. 4e also confirmed that the G-peak splitting only occurs in the ABC domains.

In Fig. 5, the molecular adsorption of iodine was utilized to induce the top-down charge injection. Among halogens with high electron affinity or oxidizing power, $I_2$ was chosen instead of $Cl_2$ or $Br_2$ since the latter two form graphite intercalation compounds (GICs).[37] It was recently shown that diffusion of $Br_2$ through graphene/$SiO_2$ interface requires a minimum equilibrium vapor pressure of 2 ~ 12 Torr.[51] To adsorb $I_2$ molecules on top surface of FLG as schematically shown in Fig. 5a avoiding their interfacial diffusion, samples were briefly exposed to $I_2$ vapor (< 0.3 Torr). As shown in the Raman spectra of Fig. 5b, the adsorption of $I_2$ also induced the G-peak splitting in ABC, but not in ABA. (See Supplementary Fig. S2 for the Raman maps of the employed sample.) It is concluded that the top-down hole injection broke the inversion symmetry of ABC and induced the mixing of the G phonon modes as explained in the bottom-up hole doping.

The G peak of ABA downshifts gradually over ~50 days following the initial rise to 1589 cm$^{-1}$ determined at ~30 min after the exposure to the $I_2$ vapor (Fig. 5c), which was attributed to the thermal desorption at room temperature. Whereas the G$^+$ peak of ABC showed a change almost identical to the G peak of ABA, it further shows that its downshift during the first 4 days is steeper than the rest, suggesting existence of multiple dopant species or complex desorption mechanisms. The G$^-$ peak frequency also exhibited the fast and slow changes but in the opposite direction. The less obvious downshift of G$^-$ in Fig. 3 can be attributed to the interference by the annealing-induced lattice compression that upshifts not only G$^+$ but also G$^-$. It is also to be noted that the molecular doping decreased the linewidths of G peaks ($\Gamma_G$) significantly as shown in Supplementary Fig. S3. The 'lifetime narrowing' can be attributed to the attenuated Landau damping caused by the downshift of the Fermi level as shown in single layer graphene (SLG).[48,52] The larger $\Gamma_G$ of doped ABA than that of doped ABC could be due to the presence of more symmetry-allowed electronic transitions[53] that lead to additional electron-phonon scattering in doped ABA. Alternatively, the larger $\Gamma_G$ of doped ABA could be due to the presence of other G peaks that could not be resolved in the spectra because of their finite linewidths.

Δω$_G$ in Fig. 5d provides an empirical estimate for the relative coverage of the adsorbed hole dopants as will be discussed quantitatively below, since the degree of the G-peak splitting is roughly proportional to the induced charge density.[32,34] It can be seen that half of the initial splitting was recovered during the early 4 days and the rest underwent a very slow recovery with a time constant of ~45 days when assumed to follow a first-order kinetics. A recent vdW density functional calculation[50] suggested that $I_2$ adsorbs on graphene at various binding sites either in an in-plane or in a vertical orientation with a maximum binding energy of ~0.50 eV at the bridge site. The calculation also confirmed the electron transfer from graphene to $I_2$ adsorbates downshifting the Fermi level below the Dirac point. The slow-decay in Fig. 5d can be attributed to $I_2$ molecules in the strongest interaction with graphene whereas the fast-decaying components to those accommodated at less favorable binding sites on graphene or in the second molecular layer.[54,55] The long-lingering adsorbates could be readily removed by rinsing with methanol as shown by the recovery of G and 2D peaks (see Supplementary Fig. S4).

To induce bottom-up and top-down doping simultaneously, samples were monitored in situ in an optical cell containing a small piece of $I_2$ crystal, which gradually sublimed to reach the vapor pressure of $I_2$ of ~0.3 Torr at room temperature. According to Jung et al.,[37] iodine molecules intercalate through the graphene-silica interface at a partial pressure of ~0.1 Torr and form two saturated dopant layers underneath and on top of 2L graphene, respectively. As shown in Fig. 6, the G peak of AB layer split into G$^-$ and G$^+$ because of the top-down hole doping (Fig. 4) for short exposure time ($t \leq 1.5$ hours). For $t \geq 2.0$ hours, however, G$^+$ further upshifted to 1604 cm$^{-1}$ whereas G$^-$ disappeared, which indicates the double-sided charge doping by the adsorbed and the intercalated $I_2$ layers as shown by the previous studies.[37,56] The G peak of ABC 3L also exhibited the splitting into G$^-$ and G$^+$ for short exposure ($t \leq 1.5$ hours) and the upshift of G$^+$ to 1600 cm$^{-1}$ at the expense of G$^-$ for extended exposure ($t \geq 2.0$ hours). We note that a new peak (denoted as G$^o$ for its non-dispersive character regardless of $t$ unlike the other G-related peaks) appeared at 1586 cm$^{-1}$ when ABC was double-side doped ($t \geq 1.5$ hours). In a similar measurement, the G peak of ABA 3L showed a monotonous upshift to 1600 cm$^{-1}$ without the splitting and G$^o$ also appeared at 1586 cm$^{-1}$ ($t \geq 1.5$ hours).

The asymmetry-induced G peak splitting of AB (ABC) layers is attributed to the anticrossing coupling between the $E_g$ and the $E_u$ modes, which are essentially mixed into the two Raman-active superposed modes.[32] Yan et al. confirmed the non-crossing behavior in the gated AB layers.[34] Based on the phonon mixing model, the hole density induced by the bottom-up doping could be estimated using the degree of G peak splitting ($\Delta\omega_G$) (see Supplementary Fig. S5). The effective hole density in the AB layers annealed at 400 ~ 700 °C lies in the range of 0.9 ~ 1.5x10$^{13}$ /cm$^2$. These values are in good agreement with those reported for 1L graphene annealed at similar temperatures.[19,43,44] Since the doping is believed to be driven by the interfacial redox reaction,[44-46] the induced charge density will be limited by the density of the redox couples at the interface and thus constant regardless of the thickness of graphene, assuming that the small difference[57] in their work functions can be disregarded. The mixing model further explains how the G peak intensity bifurcates when split. According to the theoretical calculation by Ando et al.,[32] the intensity of G$^-$ (G$^+$) decreases (increases) with increasing charge density. As shown in Supplementary Fig. S5b, the fractional intensities of G$^-$ and G$^+$ for the AB and ABC layers agree well with the theoretical prediction.[32,34] However, their linewidths did not follow the theoretical prediction, which can be attributed to the inhomogeneous broadening due to annealing[43] (see Supplementary Fig. S5c).

One may argue that G$^-$ and G$^+$ originate respectively from the unevenly doped upper and bottom layers in the case of AB layers. Although the so called local Raman model has been successfully used in interpreting the G peak splitting of GICs, it failed to explain the significant intensity difference between the split G peaks of FLG samples.[31,37] In contrast, the phonon mixing model well describes how the spectral intensity is transferred between the superposed modes as the degree of asymmetry increases.[32,34] Alternatively, one may consider the fact that the Raman-inactive $E_u$ mode belongs to the Raman-active $E$ representation when the inversion symmetry is lifted.[31,58] Whereas this explains the G peak splitting or the emergence of the second G peak (G$^-$), the mode is not likely to constitute the eigen states of the symmetry-broken AB layers.[34] In addition, the observation of the decreasing intensity of the G$^-$ peak corresponding to the $E_u$ mode with increasing degree of asymmetry contradicts the assignment.[34]

Charge screening determines how charges are distributed across the sub-layers and thus how much each layer contributes to transporting electrical currents in FLG devices.[59,60] Various unwanted influence of charge defects in substrates is also dictated by the degree of screening.[61] Due to the vanishing electronic density at the Dirac point, the screening in graphene is generally very weak but exhibits unconventional temperature dependence.[62] In FLG[60] and GICs,[63] essentially electrostatically coupled stacks of graphene,[48] the screening is also a complex function of charge density and temperature as exemplified by experimental screening lengths scattered over an order of magnitude.[60] The G splitting observed in the chemically doped AB and ABC layers confirms that the charge distribution is significantly uneven among the sub-layers suggesting that the charge screening length is on the order of their interplanar distance. This is consistent with the fact that charge transfer from the intercalate layer is largely localized within the bounding graphene layers of GICs.[63,64] Nevertheless, the observed splitting of the single-side doped FLG cannot be interpreted in the local Raman model[65] proposed for GICs.

Our study also revealed that G⁻ disappears and a new G-related peak, Gº, appears when one-side-doped ABC 3L is further doped from the other side. Note that the G⁻ peaks in the double-side doped 3L reported by Jung et al.[37] and Zhao et al.[55] in fact correspond to Gº, distinct from G⁻ of unidirectionally doped 3L. Our study clearly showed that Gº is seen in both ABC and ABA 3L, which was not resolved in the previous studies.[37,56] It is also notable that Gº remained at ~1586 cm$^{-1}$ while G upshifted from 1592 cm$^{-1}$ to 1600 cm$^{-1}$. We speculate that Gº and G (at 1600 cm$^{-1}$) originate respectively from the weakly doped interior layer and the strongly doped bounding layers in the local Raman model.[37,66,67] Note that the G-related Raman spectra of double-side doped nL graphene in the current work and previous studies[37,55,67] are in good agreement with those from stage-n GIC intercalated with $FeCl_3$.[66] In fact, the double-side doped nL can be considered as a building block of stage-n GICs. However, the intensity ratio of $I$(G)/$I$(Gº) for 3L in Fig. 6 is ~4, twice larger than that of the stage-3 GICs[66] or what is expected from the geometrical consideration. The higher $I$(G)/$I$(Gº) suggests that charge penetrates beyond the outermost layer of 3L. Crowther et al.[67] concluded that the two outermost layers at each side of double-side doped FLG (n > 4) are effectively doped. The longer charge screening length of FLG compared to graphite may be due to the fact that dopant layers contact graphene layers at only one side thus experiencing less screening.

Manipulating the symmetry of 2-dimensional crystals will be useful not only for the fundamental science but also for creating new applications. This study demonstrated that the crystalline symmetry of FLG can be manipulated by the simple chemical treatments that inject extra charge carriers selectively from the top or the bottom side of the graphene sheets. Whereas the electrostatic gating[48,52] has been widely used for fine and rapid tuning of the charge density, the chemical method can be complementary in that it does not require delicate electrical connection thus can be used for systems of arbitrary dimensions on any substrates. Moreover, molecular adsorption can readily lead to a large degree of charge transfer inducing an optical gap of ~2 eV.[51] It should be also possible to enhance the density tuning range further via a push-pull type double-sided chemical doping.[68] The molecular desorption at the ambient temperature may be suppressed by encapsulating the system in a sandwich structure.[69] Our study also showed that the modified crystalline symmetry can be readily recovered by removing the dopants by the molecular replacement or rinsing.

Due to the dominant role of molecular interactions with the confining walls, understanding the unique molecular behavior in nanoscopic confined space bears significant implication in diverse fields such as biology, geology, meteorology and nanotechnology.[70-72] FLG sheets and the underlying substrates form ideal two dimensional space with a van der Waals gap of a few angstroms, where FLG sheets serve not only as confining flexible walls but also as transparent spectral windows.[40] Our study shows that FLG sheets can also be used as a spectroscopic indicator which is sensitive to molecular behaviors occurring at the interface.[40] Similar approaches can be applied to other various 2-dimensional crystals, which will bring us deeper understanding of the nanoscopic world. We also note that the interfacial molecular diffusion used for the replacement can be further exploited for purposeful chemical doping in various 2-dimensional material systems. Pre-designed surface functional groups of the substrates can be used to not only dope graphene,[73,74] but also control the process of the interfacial diffusion.

In summary, we have demonstrated that the crystalline symmetry of AB and ABC FLG can be modulated by chemical charge doping. Extra charge carriers could be injected into FLG in the top-down and bottom-up direction, respectively, by molecular adsorption of $I_2$ and thermally inducing interfacial hole dopants. The breakdown of the inversion symmetry could be monitored by the G peak splitting occurring as a result of the mixing of the G phonons, even-parity $E_g$ and odd-parity $E_u$ modes. Simple rinsing and interfacial molecular replacement were shown to recover the broken symmetry. We also showed that ABA trilayers lacking inversion symmetry does not undergo G-peak splitting when chemically doped. By extending exposure to $I_2$ vapor, FLG could also be doped simultaneously in both directions. In addition to the stiffened G peak, the double-side doped 3L exhibited the $G^o$ peak, distinct from the $G^-$ peak of the single-side doped FLG. Modification of crystalline symmetry through the reversible and direction-controlled chemical doping demonstrated in this study can also be useful in modifying the electronic and vibrational transitions of other low dimensional crystals.

**Methods**

**Preparation of samples.** FLG samples were prepared in the ambient conditions using the micromechanical exfoliation method developed by Geim's group.[1] Briefly, a piece of kish graphite (Covalent Materials Inc.) was exfoliated into multiple thinner flakes using adhesive tapes and mechanically transferred on Si wafer substrates covered with 285 nm thick $SiO_2$ layers. After locating very thin flakes spanning more than 10 μm across using an optical microscope, their number of layers and degree of crystallinity were determined using Raman spectroscopy.[75-77]

**Raman spectroscopy.** The details of the home-built micro Raman setup can be found elsewhere.[43,50] Briefly, the 514.5 nm output (≤ 1.3 mW) of an Ar ion laser was focused at a diffraction-limited spot (< 1 μm) on the sample plane using an objective lens (40X, numerical aperture = 0.60). The backscattered

Raman signal collected by the same objective lens was fed into a spectrograph (focal length = 300 mm) equipped with a liquid nitrogen-cooled CCD detector. The spectral resolution defined by the linewidth of the Rayleigh line was 3.0 cm$^{-1}$ and spectral accuracy was better than 1 cm$^{-1}$.[43] Raman spectra in Fig. 4 were obtained with 632.8 nm HeNe laser output ($\leq$ 3.3 mW), where the spectral resolution was 3.5 cm$^{-1}$.

**Bottom-up chemical doping and its reversal.** The bottom-up hole doping was induced by the annealing-induced hole doping.[19,40,44] After characterized in their pristine state by Raman spectroscopy, samples were annealed in a quartz tube furnace in a vacuum (< 3 mTorr). The temperature was linearly ramped to a target ($T_{anneal}$) in 30 min and maintained for 2 hours followed by spontaneous cooling down to ~23 °C. When exposed to the air after the annealing, ambient oxygen molecules[44] intercalate through the graphene/SiO$_2$ interface[40] and undergo a redox reaction that injects holes into the graphene.[45,46] The hole density remained stable in the ambient conditions for more than several months. The degree of charge density determined by Raman spectroscopy[43] could be varied up to ~2x10$^{13}$ /cm$^2$ for 1L by raising $T_{anneal}$ to 600 °C.[19] To monitor the reversal of the hole doping, Raman spectra were obtained in situ while samples mounted in an optical liquid cell were submerged in distilled water according to D. Lee et al.'s work.[40] It was shown that intercalation of ultrathin water layer undopes graphene almost completely.

**Top-down chemical doping and its reversal.** To induce hole doping in a top-down manner, FLG samples were briefly exposed to I$_2$ vapor from a small glass vial containing I$_2$ crystals in the ambient conditions. Since halogen molecules may intercalate through graphene/SiO$_2$ interface for prolonged exposure,[51] the exposure was minimized to a level that avoids the double-sided doping as explained below. Adsorbed I$_2$ is known to inject high density of electrical holes.[37] The adsorbed iodine species could be readily removed by rinsing with methanol.

**Double-sided chemical doping.** To induce double-sided hole doping, FLG samples were placed in an optical cell which contained a small piece of I$_2$ crystal. As increasing the exposure time (*t*) defined as the time that lapsed since the encapsulation, the partial pressure of I$_2$ vapor gradually increases to ~0.3 Torr or its vapor pressure[78] at room temperature. According to Jung et al.,[37] the double-sided doping was confirmed by the disappearance of G$^-$ and upshift of G to 1604 cm$^{-1}$ for AB 2L and was found to be obtained within a few hours of exposure (Fig. 6).

**Author contributions**

S.R. proposed and supervised the project. K.P. performed the experiments and analysed the data. S.R. and K.P. wrote the manuscript.

**Acknowledgments**

This work was supported by the Center for Advanced Soft-Electronics funded by the Ministry of Science, ICT and Future Planning as Global Frontier Project (CASE-2014M3A6A5060934) and also by the National Research Foundation of Korea (NRF-2012R1A1A2043136).

**Additional information**

Supplementary information accompanies this paper at http://www.nature.com/scientificreports

Competing financial interests: The authors declare no competing financial interests.

**Figure captions**

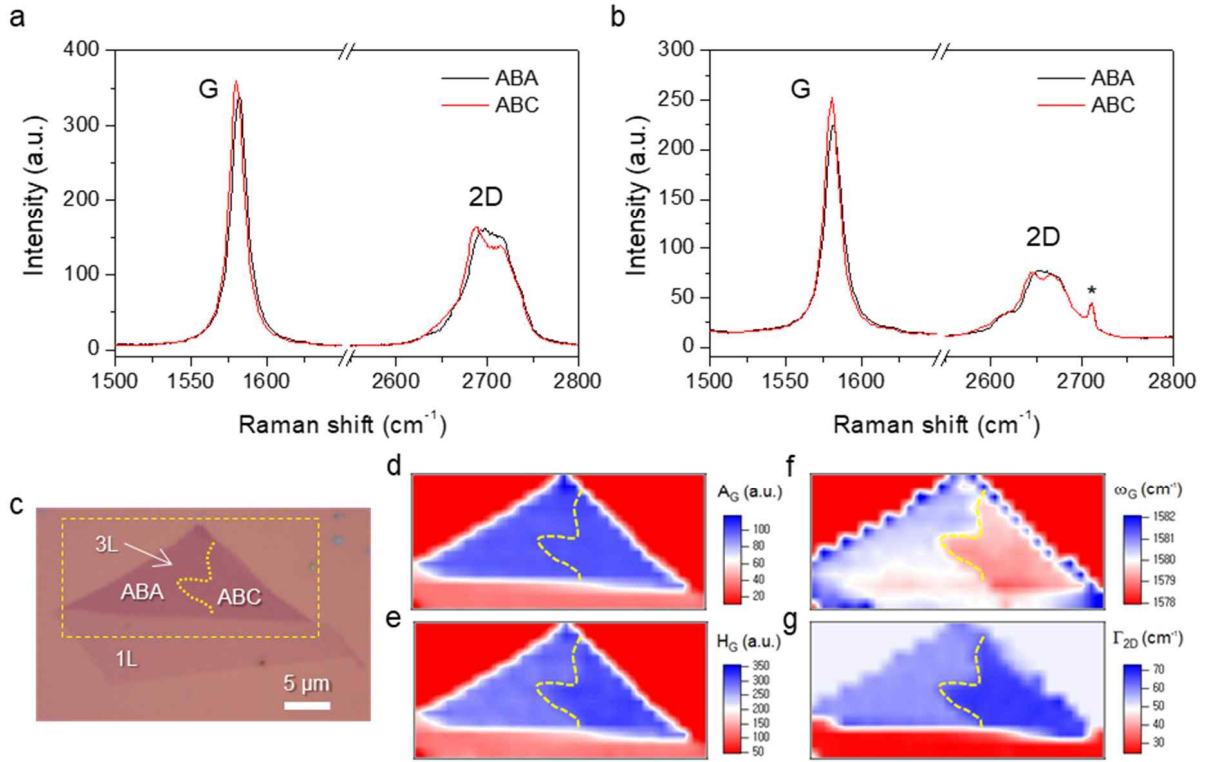

Figure 1. Raman characterization of pristine trilayer (3L) with ABA and ABC domains. (a) Raman spectra of ABA and ABC trilayers obtained with excitation wavelength of 514 nm. (b) Raman spectra of ABA and ABC trilayers obtained with excitation wavelength of 633 nm. The asterisked peaks at ~2710 cm$^{-1}$ originated from the excitation laser. (c) Optical micrograph of 3L graphene with attached 1L. (d~g) The Raman maps obtained with excitation wavelength of 514 nm from the dashed rectangle in (c): (d) the peak area of G ($A_G$), (e) the peak height of G ($H_G$), the peak frequency of G ($\omega_G$), (e) the linewidth of 2D ($\Gamma_{2D}$). Whereas the step size of mapping was 1 micron, the map images were refined by bilinear interpolation. The dotted ABA-ABC boundary in (c~g) was determined from the $\Gamma_{2D}$-Raman map shown in (g).

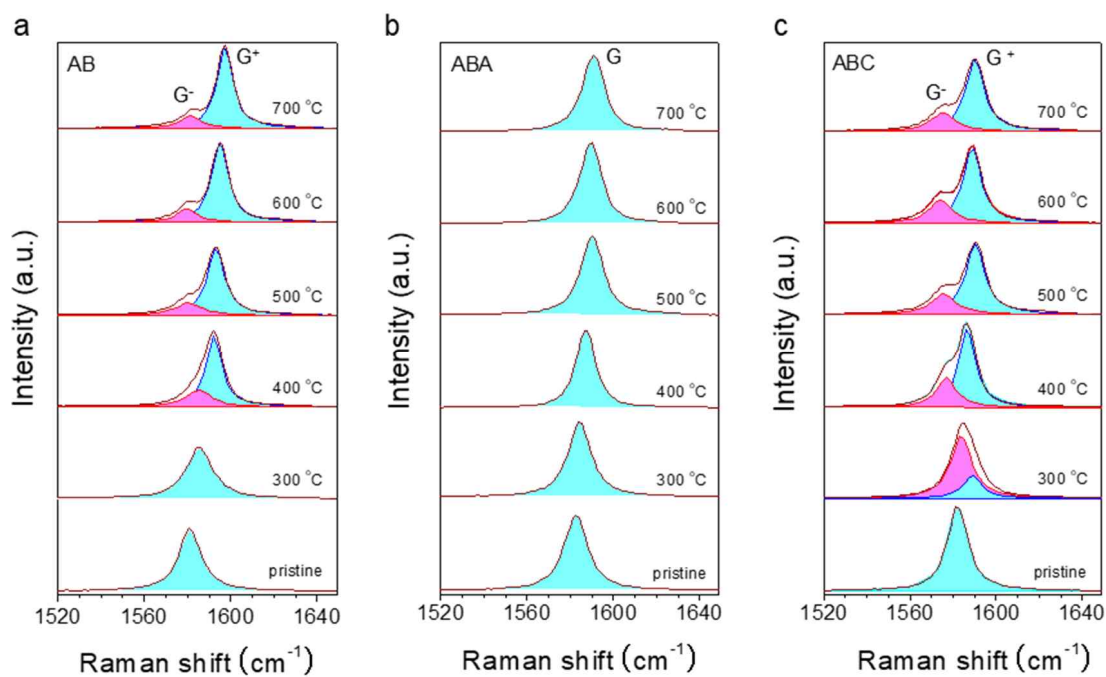

Figure 2. Effects of the bottom-up hole injection in FLG/SiO$_2$/Si substrates by thermal annealing: Raman spectra of (a) AB, (b) ABA and (c) ABC FLG, annealed in a vacuum at elevated temperatures ($T_{anneal}$). Upon annealing, G peak of AB and ABC splits into G$^-$ and G$^+$ unlike that of ABA.

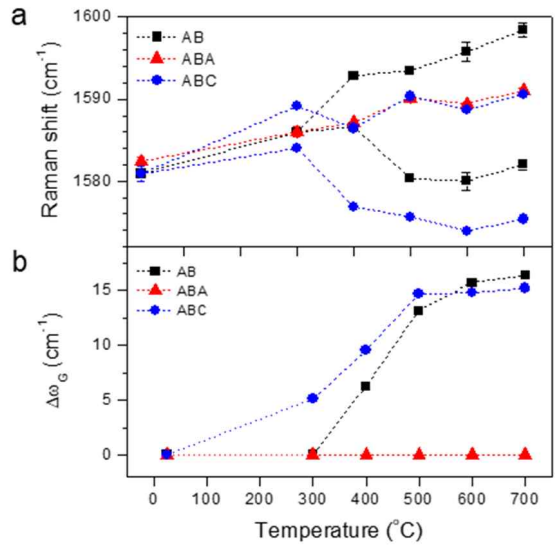

Figure 3. G-peak splitting of AB and ABC due to the bottom-up hole injection. (a) The frequency of G, G$^-$ and G$^+$ as a function of T$_{anneal}$. (b) The frequency difference ($\Delta\omega_G$) between G$^-$ and G$^+$ as a function of T$_{anneal}$.

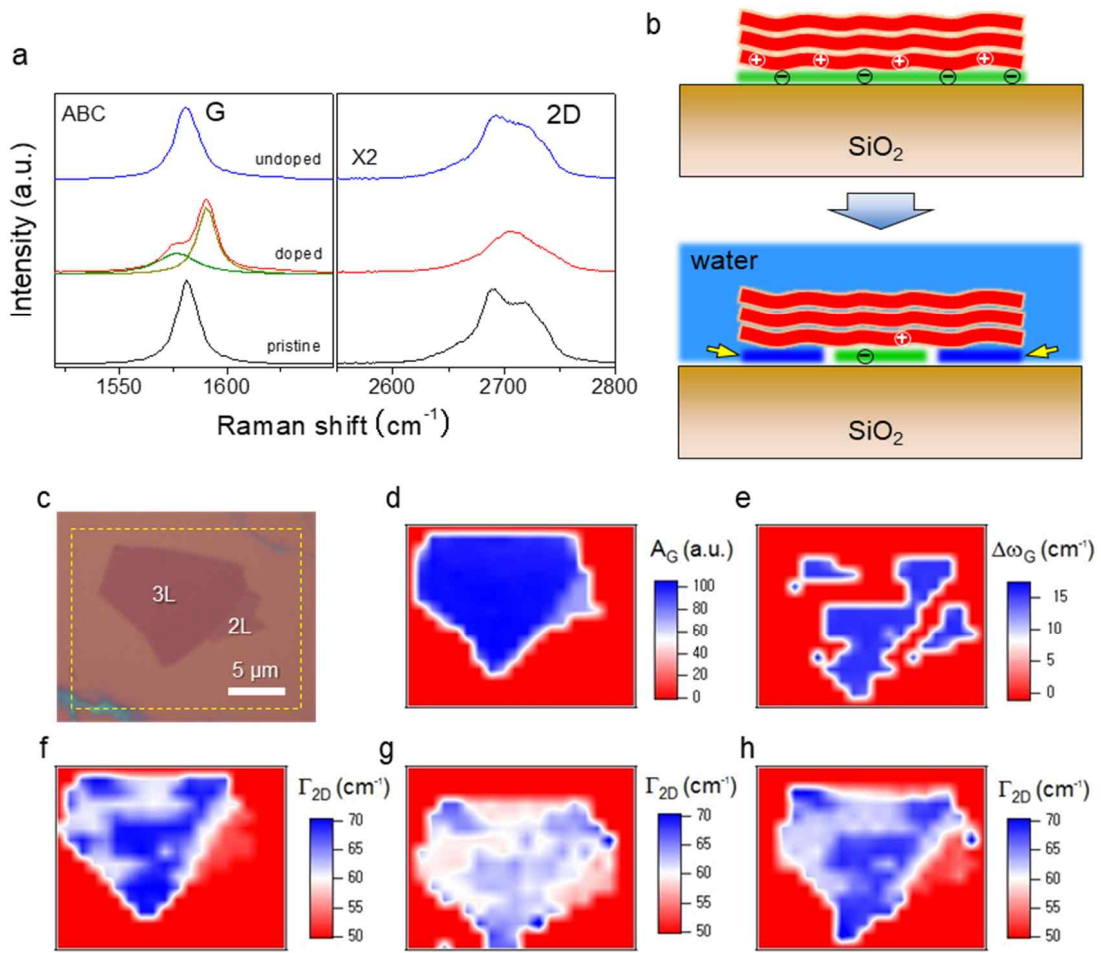

Figure 4. Reversal of the bottom-up hole injection via water intercalation. (a) Raman spectra of ABC domain in its pristine, doped and undoped states. $T_{anneal}$ was 500 °C and $t_{water}$ was 7 days. (b) Schematic diagrams of water intercalation from the edge to the center through the graphene-silica interface. The green and dark blue layers represent the hole dopants and the intercalated water layer, respectively. (c) Optical micrograph of the pristine FLG sample. (d) $A_G$-map of the pristine sample. The map was obtained from the dashed rectangle in (c). (e) $\Delta\omega_G$-map of the doped sample. (f, g, h) $\Gamma_{2D}$-maps obtained in its pristine, doped and undoped states, respectively.

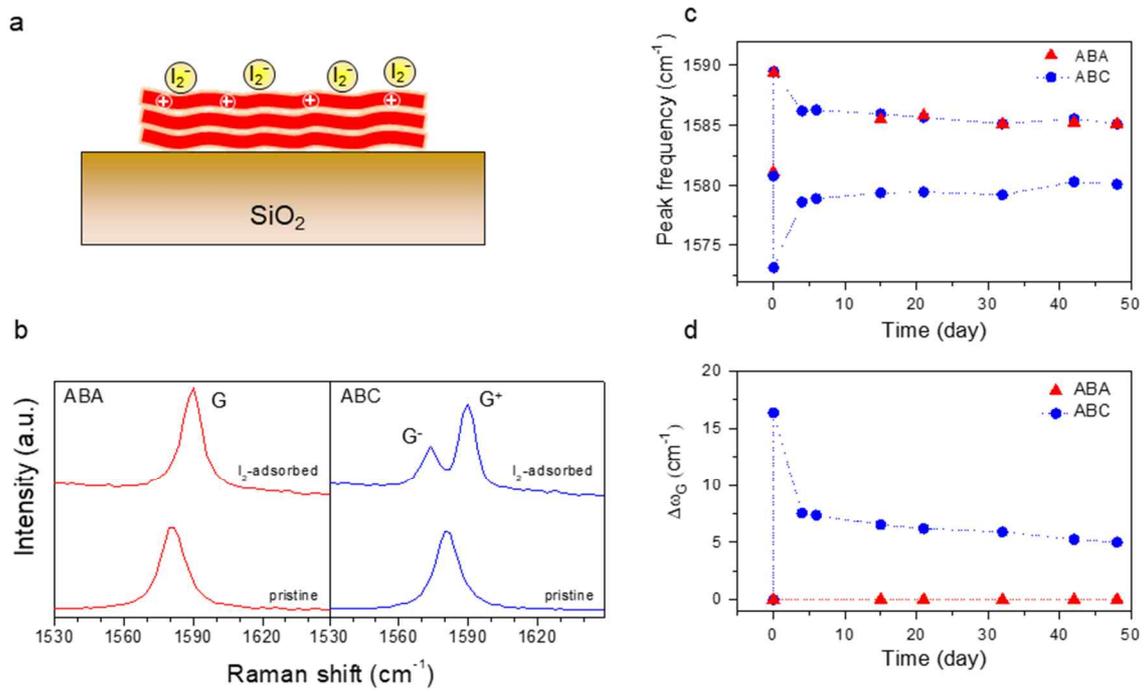

Figure 5. Effects of the top-down hole injection in FLG/SiO$_2$/Si substrates by adsorption of I$_2$. (a) Schematic diagram of the top-down charge doping in 3L with iodine molecules. (b) Raman spectra of ABA and ABC in their pristine and I$_2$-adsorbed states. (c) The frequency of G, G$^-$ and G$^+$ as a function of elapsed time since the adsorption of I$_2$. (d) The peak frequency difference ($\Delta\omega_G$) between G$^-$ and G$^+$ as a function of the elapsed time.

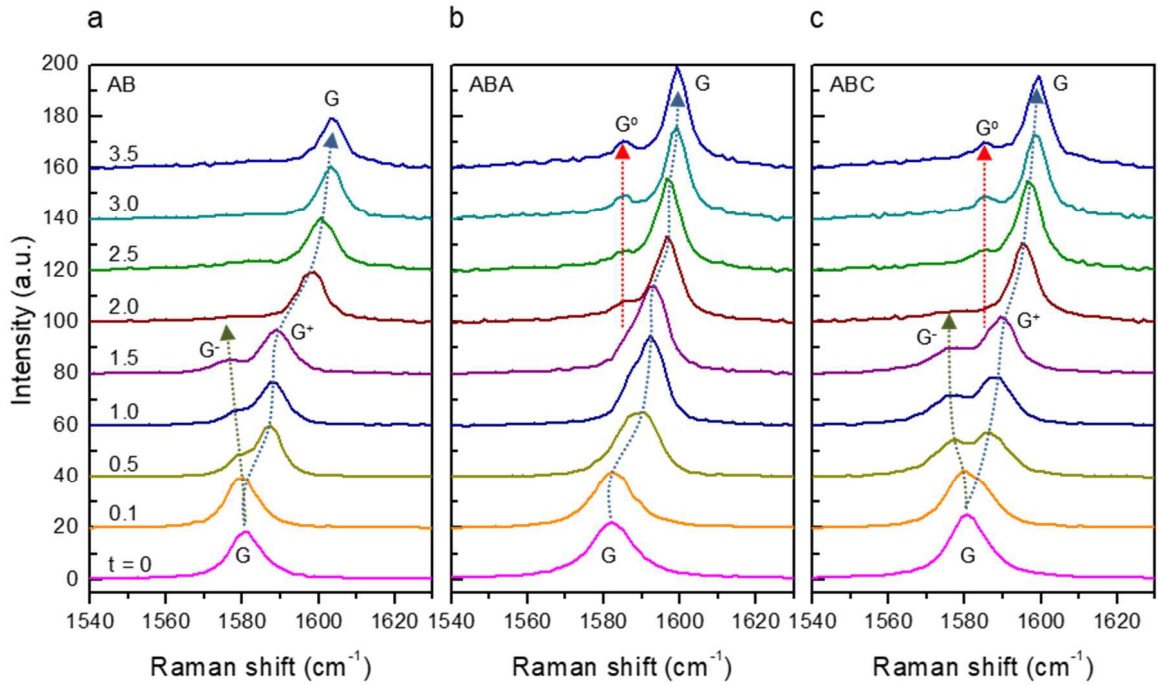

Figure 6. Single-sided vs double-sided hole injection in FLG/SiO$_2$/Si substrates by I$_2$. The Raman spectra of AB (a), ABA (b), and ABC (c) obtained in an optical cell as a function of the exposure time (*t*) to I$_2$ vapor. The spectra were vertically offset for clarity after the broad fluorescence from iodine species was subtracted from each spectrum.